\documentclass[twocolumn,superscriptaddress,nofootnoteinbib]{revtex4}

\usepackage{amsfonts}
\usepackage{amsmath}
\usepackage{graphicx}
\usepackage{amssymb}
\usepackage[british]{babel}
\usepackage{color}
\usepackage[T1]{fontenc}
\usepackage{bm}
\usepackage{mathrsfs}
\usepackage{marvosym}
\usepackage[colorlinks=true,linkcolor=blue,citecolor=blue]{hyperref}

\begin{document}

\title{Towards a robust criterion of anomalous diffusion}

\author{Vittoria Sposini}
\affiliation{Faculty of Physics, University of Vienna, Kolingasse 14-16, 
A-1090 Vienna, Austria}
\author{Diego Krapf}
\affiliation{Department of Electrical and Computer Engineering, Colorado
State University, Fort Collins, CO 80523, USA}
\affiliation{School of Biomedical Engineering, Colorado State University,
Fort Collins, CO 80523, USA}
\author{Enzo Marinari}
\affiliation{Dipartimento di Fisica, Sapienza Universit{\`a} di Roma,
P.le A. Moro 2, I-00185 Roma, Italy}
\affiliation{INFN, Sezione di Roma 1 and Nanotech-CNR, UOS di Roma,
P.le A. Moro 2, I-00185 Roma, Italy}
\author{Raimon Sunyer}
\affiliation{Unitat de Biof\'{i}sica i Bioenginyeria, Departament de
Biomedicina Facultat de Medicina, Universitat de Barcelona, Diagonal
647, 08028 Barcelona, Spain}
\author{Felix Ritort}
\affiliation{Small Biosystems Lab, Condensed Matter Physics Department, Facultat
de F\'{i}sica, Universitat de Barcelona, Diagonal 647, 08028 Barcelona, Spain}
\author{Fereydoon Taheri}
\affiliation{Institute for Molecular Systems Engineering and Advanced
Materials (IMSEAM), Heidelberg University, INF 225, 69120 Heidelberg, Germany}
\author{Christine Selhuber-Unkel}
\affiliation{Institute for Molecular Systems Engineering and Advanced
Materials (IMSEAM), Heidelberg University, INF 225, 69120 Heidelberg, Germany}
\author{Rebecca Benelli}
\affiliation{Experimental Physics I, University of Bayreuth, D-95440
Bayreuth, Germany}
\author{Matthias Weiss}
\affiliation{Experimental Physics I, University of Bayreuth, D-95440
Bayreuth, Germany}
\author{Ralf Metzler$^\ast$}
\affiliation{Institute of Physics and Astronomy, University of Potsdam,
Karl-Liebknecht-Str 24/25, 14476 Potsdam-Golm, Germany}
\email{rmetzler@uni-potsdam.de}
\affiliation{Asia Pacific Centre for Theoretical Physics, Pohang 37673, 
Republic of Korea}
\author{Gleb Oshanin}
\affiliation{Sorbonne Universit\'e, CNRS, Laboratoire de Physique
Th\'{e}orique de la Mati\`{e}re Condens\'{e}e (UMR 7600), 4 Place
Jussieu, 75252 Paris Cedex 05, France}
\email{gleb.oshanin@sorbonne-universite.fr}

\begin{abstract}
Anomalous-diffusion, the departure of the spreading dynamics of diffusing
particles from the traditional law of Brownian-motion, is a signature
feature of a large number of complex soft-matter and biological
systems. Anomalous-diffusion emerges due to a variety of physical
mechanisms, e.g., trapping interactions or the viscoelasticity of the
environment. However, sometimes systems dynamics are erroneously claimed
to be anomalous, despite the fact that the true motion is Brownian---or
vice versa. This ambiguity in establishing whether the dynamics as normal
or anomalous can have far-reaching consequences, e.g., in predictions
for reaction- or relaxation-laws. Demonstrating that a system exhibits
normal- or anomalous-diffusion is highly desirable for a vast host of
applications. Here, we present a criterion for anomalous-diffusion based on
the method of power-spectral analysis of single trajectories. The robustness
of this criterion is studied for trajectories of fractional-Brownian-motion,
a ubiquitous stochastic process for the description of anomalous-diffusion,
in the presence of two types of measurement errors. In particular, we find that
our criterion is very robust for subdiffusion. Various tests on surrogate data
in absence or presence of additional positional noise demonstrate the efficacy
of this method in practical contexts. Finally, we provide a proof-of-concept
based on diverse experiments exhibiting both normal and anomalous-diffusion.
\end{abstract}

\maketitle

\section{Introduction}

The exploration of the dynamic properties of complex systems has been
massively boosted by modern microscopic techniques allowing single-particle
tracking (SPT) of micron- and submicron-sized tracers or even single molecules.
SPT is routinely used to probe the local properties of materials and even live
biological cells and tissue by passive and active microrheology \cite{zia}.
SPT by now is a key tool to interrogate the structure--function relationships
in biophysical applications \cite{shen,manzo}, and it plays a central role
in uncovering thermal and energy-fuelled intracellular transport of tracer
particles or single molecules in biological cells and tissues \cite{mogre}. 
SPT is thus at the heart of the new emerging era of quantitative life
sciences \cite{golding,pt,pccp,manzo,norregaard,pt1,seisenhuber,he,platani,
PSD_Diego,etoc,song,krzysiek,ramm,normanno,wuite,rienzo,golan,shen,moerner,
granick,mogre,tabei,garini,weber,lene,thapa2019, ohad, Nathan_Science,
Erickson2015}. 
Specifically, SPT unveiled different intracellular motion patterns of virus 
particles \cite{seisenhuber}, Cajal bodies \cite{platani}, molecular motor-driven 
transport \cite{song,granick,ramm}, the motion of telomeres \cite{garini}, 
green fluorescent proteins \cite{golan}, DNA-binding proteins 
\cite{normanno,wuite}, mRNA molecules \cite{moerner,golding,weber}, 
membrane proteins \cite{rienzo, PSD_Diego, he}, or endogenous granules 
\cite{tabei,lene} and vesicles \cite{thapa2019}. 
SPT has also revealed protein interactions \cite{krzysiek} as well as key
details of submicron tracer motion in mammalian cells \cite{etoc} and 
in the movement ecology of larger animals \cite{Nathan_Science, Vilk_PRX}.

In contrast to pre-averaged data such as those obtained from 
fluorescence correlation spectroscopy (FCS) or fluorescence recovery after 
photobleaching (FRAP), SPT provides high resolution "unprocessed" data: 
As raw data of test particle trajectories, SPT offers the best possible basis 
for statistical analysis \cite{braeuchle}. Indeed, a large toolbox of methods 
are available for analysing position time series $X_t$. Frequently, these 
data have revealed that particles exhibit an anomalous diffusion behaviour, 
defined by the power-law dependence $\langle X^2_t\rangle\simeq t^{\alpha}$ 
of the mean squared displacement (MSD), where the angular brackets
denote averaging over different realisations of recorded trajectories. The 
anomalous diffusion exponent $\alpha$ is commonly used to tell whether 
normal (Brownian, $\alpha=1$) or anomalous diffusion ($\alpha\ne1$) 
is observed. Here, the regimes of sub- and superdiffusion correspond to 
$0<\alpha <1$ and $\alpha>1$, respectively. Distinguishing normal 
from anomalous diffusion is vital for predicting various characteristics 
of the systems under investigation---e.g., the diffusion-control of
molecular reactions or the relaxation dynamics after manipulating the
system---eventually allowing us to understand the actual physical
mechanisms underlying the observations.

Fitting the scaling exponent $\alpha$ to finite measured time series is
known to be a major challenge. For instance, contamination of the true
trajectories by measurement noise was shown to lead to the erroneous
conclusion to observe anomalous diffusion ($\alpha\neq1$) \cite{kaes}. 
There exist methods to alleviate this problem, e.g., the mean-maximal 
excursion statistics \cite{vincent}. Moreover, Bayesian-maximum likelihood 
methods \cite{samu,samusoft,michael}, deep learning strategies \cite{gorka,
yael,janusz,januszentro,andi}, or feature-based methods \cite{janusz,
januszentro,amanda,jakub} also provide best estimates for $\alpha$.
However, all these methods have their shortcomings. Quite severely, 
$\alpha$ values (along with $D$ values \cite{boyer,Flyvbjerg}) retrieved 
from fitting the MSD $\langle X^2_t\rangle$ will vary from one trajectory 
to the next due to finite statistics within single trajectories. Indeed,
trajectories from real experiments display different regimes with different 
scaling exponents \cite{samu,samusoft,yaelprr}, or due to the spatial
heterogeneity of the environment \cite{lanoiselee, nickelife}. This leads to strong 
variability in the scaling exponents measured for different trajectories and
renders predictions based on such fitting procedures even less accurate. 
Moreover, for realistic situations with $\alpha$ values closer to the 
Brownian value $\alpha=1$ it becomes increasingly difficult to distinguish 
anomalous from Brownian motion \cite{januszentro}. This latter caveat 
is further exacerbated when considering the unavoidable experimental 
sources of error in many SPT setups, especially when based on 
fluorescence microscopy methods.

Likely the most notorious source of uncertainty is the static localisation
error that arises from the finite number of fluorescence photons 
garnered during an image of the particle from which its position 
will be retrieved by elaborate tracking schemes: Each of these photons 
is emitted from a point-like source in the sample (the emitting 
fluorophore) and will hence be captured on a locus of the camera 
sensor according to the microscope's point-spread function (PSF). 
In other words, individual photons are stochastically recorded on the 
camera sensor, with a distribution of positions around the actual particle 
location determined by the PSF \cite{ober2004localization}. As a 
consequence, recording only a few photons will yield a poor estimate 
for the actual particle position, and the characteristic deviation is 
determined by the standard error. The latter scales with the inverse root 
of the number of photons, i.e., for a large number of photons the static 
localisation error can be as small as a few nanometres \cite{yildiz2005fluorescence}. 
However, for very large numbers of photons another perturbation 
becomes visible, the so-called dynamic localisation error: Recording an 
image to determine a particle's position takes a finite time during which 
particles are constantly on the move. As a result, many different positions 
are visited during the acquisition of a single image and only the temporal 
mean of these is retrieved from the acquired image as the, apparent, 
particle position. This dynamic localisation error effectively adds a negative 
offset to $\langle X^2_t\rangle$ whereas the static localisation error adds 
a positive offset \cite{Doyle,backlund2015chromosomal}. Both sources 
of error will therefore perturb the analysis of the scaling behaviour on short 
time scales for which experimental trajectories typically yield best statistics. 
Thus, determining the value of $\alpha$ and deciding whether diffusional 
anomalies are present is indeed a major challenge.

The aim of this article is the analysis of a robust and easy to 
implement method that allows one to decide on the type 
and significance of an apparent anomaly, without being spoiled
by localisation errors. We concentrate here on the situation when $X_t$ 
belongs to a wide, experimentally-relevant class of anomalous 
diffusions---the so-called fractional Brownian motion (FBM) \cite{Mandelbrot}. 
Note, however, that the methodology we develop here will be amenable
to generalisation to any anomalous--diffusion process.
FBM is a Gaussian stochastic process characterised by a zero mean value 
and the covariance function
\begin{equation}
\langle X_{t_1}X_{t_2}\rangle=D\left(t_1^{2H}+t_2^{2H}-\left|t_1-t_2
\right|^{2H}\right),
\label{covFBM}
\end{equation}
where $D$ is a proportionality factor with physical units of $\mathrm{length}
^2/\mathrm{time}^{2H}$ commonly referred to as generalised diffusion 
coefficient and $H\in(0,1)$ is the traditionally used Hurst index, such that 
the anomalous diffusion exponent is $\alpha=2H$. FBM thus describes 
a process that can be subdiffusive ($H<1/2$), diffusive ($H=1/2$), or 
super-diffusive $(H>1/2)$. From a physical point of view, FBM is well 
suited to model diffusion in viscoelastic media \cite{fBM_Weiss,lene,lene1,
weber,pvar}, but it also governs observed motion patterns in movement 
ecology \cite{ohad}, the density profiles of serotonergic brain fibres
\cite{skirmantas}, or roughness in financial data \cite{beran}.

As mentioned before, localisation errors are usually divided into two kind
of contributions \cite{berglund, Doyle}: the static error, due to intrinsic
properties of the experimental setup, and the dynamic one, due to the
finite time needed for data acquisition, i.e., the exposure time. From a
mathematical point of view the former is generally treated as an independent
additive noise source whereas the latter is defined via temporal integration
over the finite exposure time. The effect of measurement error in SPT has
been investigated mainly by focusing on the MSD \cite{Doyle,michalet,Weiss2019}.
Few results are also present in literature concerning correlation functions
and power spectra \cite{Doyle,backlund2015chromosomal}, but with a small
range of applicability. Spectral analysis of stochastic processes can be very
helpful in their characterisation, however the power spectrum, according
to the text-book definition, is a property that relies on the measurement
time going to infinity and on a very large statistical ensemble. Both of these
assumptions are typically not met when dealing with state-of-the-art SPT data,
and this is why the spectral analysis of individual trajectories was only recently 
introduced \cite{PSD_BM}. The study of single-trajectory spectral densities has 
been carried out for different stochastic processes \cite{PSD_fBM,PSD_SBM,
PSD_DD,PSD_AS,PSD_Diego,PSD_SC, Vilk_JPhysA} and is based on the 
study of the random variable 
\cite{PSD_BM, PSD_fBM}
\begin{eqnarray}
\label{S}
\nonumber
S(f,T)&=&\frac{1}{T}\left|\int^T_0dte^{ift}X_t\right|^2\\
&&\hspace*{-1.2cm}=\frac{1}{T}\int^T_0dt_1\int^T_0dt_2\cos
\left(f(t_1-t_2)\right)X_{t_1}X_{t_2},
\end{eqnarray}
where $f$ is the frequency, $T$ is the finite observation time, and $X_t$
is an individual realisation of a given stochastic process. We will refer
to this quantity as the single-trajectory power spectral density (PSD).

When the parental process $X_t$ is Gaussian, the probability density 
function (PDF) of the random variable $S(f,T)$ is, likewise, entirely 
defined by its first moment and variance
\begin{eqnarray}
\mu(f,T)&=&\left< S(f,T)\right>,\\
\sigma^2(f,T)&=&\left< S^2(f,T)\right>-\left< S(f,T)\right>^2.
\end{eqnarray}
Note that by taking the limit $\lim_{T\to\infty}\mu(f,T)=\mu(f)$, we recover
the standard definition of the power spectrum. Interestingly it was 
shown that the characteristic trend of the ensemble averaged power 
spectrum $\mu(f,T)$ of several stochastic processes at high frequencies 
can be inferred already from a single-trajectory PSD \cite{PSD_BM,PSD_fBM,
PSD_SBM,PSD_DD,PSD_AS,PSD_Diego,PSD_SC}.

A very interesting quantity to study when performing single-trajectory spectral
analyses is the so-called coefficient of variation of the associated PDF,
$P(S(f,T))$,
\begin{equation}
\gamma(f,T)=\frac{\sigma(f,T)}{\mu(f,T)}.
\label{def_gamma}
\end{equation}
In particular, when dealing with FBM-like motion the limiting value of
$\gamma(f,T)$ at high frequencies (or for a fixed frequency but for long $T$)
turns out to be distinctly different in the cases of subdiffusion ($\gamma(f,T)
\sim1$), normal diffusion ($\gamma(f,T)\sim\sqrt{5}/2$) and superdiffusion
($\gamma(f,T)\sim\sqrt{2}$) \cite{PSD_fBM}.  This quantity was
proposed in \cite{PSD_fBM} as criterion of anomalous diffusion. Here, we 
go a significant step further and study how its trend towards specific 
limiting values is affected by the experimentally unavoidable presence of 
localisation errors in tracked trajectories. In particular, we show
that: (i) this criterion is very robust for subdiffusion; (ii) in the case of 
superdiffusion the limiting value of $\gamma(f,T)$ is affected by the static 
measurement error and not by the dynamic error; (iii) for normal diffusion both 
static and dynamic errors introduce correction terms in the limiting value of 
$\gamma(f,T)$. Knowledge of these results allows a reliable determination of
the anomalous nature of measured signals.

\section{Results and Discussion}

As already mentioned above, for pure FBM trajectories it was shown
\cite{PSD_fBM} that the coefficient of variation at high frequencies reaches
the limiting values: (i) $\gamma=1$ for subdiffusion, regardless of the value
of $H$, (ii) $\gamma=\sqrt{5}/2$ for normal diffusion, and (iii) 
$\gamma=\sqrt{2}$ for superdiffusion, regardless of the value of $H$. 
Here, we analyse the case of FBM trajectories in the presence of 
localisation errors. 

\subsection{Analytical predictions}

Let us start by introducing the mathematical description of the two
localisation errors:

(i) The static error is usually modelled as an additive noise term, thus
we denote with $Y_t$ the joint stochastic process of the form
\begin{equation}
\label{Y_t}
Y_t=X_t+e_t,
\end{equation}
where $X_t$ is a pure FBM trajectory and $e_t$ is the static error due to
an imperfect measurement. In a standard fashion \cite{michalet,berglund},
we stipulate that $e_t$ is given by the stationary Ornstein-Uhlenbeck (OU)
process
\begin{equation}
e_t=\int^t_{-\infty}d\tau e^{-(t-\tau)/\tau_0}\zeta_{\tau},
\label{stat_err}
\end{equation}
where $\tau_0$ is the characteristic relaxation time and $\zeta_t$ is a
Gaussian white noise with zero mean and covariance $\overline{\zeta_t
\zeta_{t'}}=2\sigma_e^2\delta(t-t')$. Moreover, it is commonly assumed 
that $\tau_0 \ll \Delta t$, where $\Delta t$ is the temporal resolution of the 
trajectory. In other words, we suppose that each time when an instantaneous 
position $X_t$ is recorded, the latter is specified up to a random "error" with 
the distribution
\begin{equation}
P(e_t)=\sqrt{\frac{1}{4\pi\sigma^2_e\tau_0}}\exp\left(-\frac{e^2_t}{4\sigma_e
^2\tau_0}
\right),
\label{Pe}
\end{equation}
and is independent of the previous measurements. 

(ii) The dynamic error depends on the acquisition or exposure time $\tau_e$,
such that the acquired position $Y_t$ can be written as $\overline{X_t}=
(1/\tau_e)\int_0^{\tau_e}X_{t-\xi}d\xi$, meaning that we cannot resolve
the particle motion below $\tau_e$. 

Note that the parameters $\sigma_e$ and $\tau_0$ for the static error 
and $\tau_e$ for the dynamic error are characteristic of the experimental 
setup and thus they are usually known quantities when analysing SPT 
experiments.

As treating the dynamic error within the single-trajectory PSD framework is
quite tedious and involved, we limit our analytical study to the joint
process \eqref{Y_t} in which only the static error is present and leave the
dynamic error for numerical study, see "Simulations" subsection 
in Methods.

Selecting the process in \eqref{Y_t} and performing the single-trajectory
spectral analysis as described above we obtain our central result,
\begin{equation}
\label{main}
\gamma_Y^2(f,T)=\frac{\sigma_Y^2(f,T)}{\mu_Y^2(f,T)}=\frac{\sigma^2_X+
\sigma^2_\mathrm{OU}+2\mu_X\mu_\mathrm{OU}}
{\left(\mu_X+\mu_\mathrm{OU}\right)^2},
\end{equation}
expressing the coefficient of variation of the single-trajectory PSD of the
joint process $Y_t$ via the first moments and the variances of the spectral
densities of its constituents.
In particular, for the OU process under the assumption stated above, 
we have $\mu_\mathrm{OU}\sim2\sigma_e^2\tau_0^2$ and 
$\sigma^2_\mathrm{OU} \sim4\sigma_e^4\tau_0^4$ (see Supplementary Note 1). 
The results for the FBM process are dependent on $H$ and are in general 
quite involved. We report here just the asymptotic trends and refer the 
interested readers to \cite{PSD_fBM} for more details,
\begin{subequations}
\begin{eqnarray}
H<1/2:&\mu_X(f,T) &\sim \frac{2c_H D}{f^{2H+1}}, \\
H=1/2:&\mu_X(f,T)& \sim \frac{4 D}{f^2},   \\
H>1/2:&\mu_X(f,T)& \sim \frac{2 D}{f^2}T^{2H-1}, 
\end{eqnarray}
\end{subequations}
and $\sigma^2_X(f,T) \sim 4D^2 \left( \frac{c_H^4}{f^{4H+2}} + 
\frac{2c_H}{f^{2H+3}}T^{2H-1} +\frac{2}{f^4}T^{4H-2} \right)$, where 
$c_H=\Gamma(2H+1)\sin(\pi H)$, and $\Gamma(z)$ is the Gamma 
function \cite{PSD_fBM}. Note that in the case of superdiffusion for 
fixed $T$ the $1/f^2$ trend could erroneously lead us to the conclusion
of having standard diffusion. In addition, the superdiffusive result also 
shows a dependence on $T$, a clear feature of ageing, that helps 
in differentiating it from normal diffusion.

The limiting value at high frequencies of the coefficient of variation 
obtained in \eqref{main} is then given by
\begin{subequations}
\begin{eqnarray}
\label{eq:sub}
\hspace*{-0.8cm}H<1/2:&&\gamma_Y^2(f,T)\sim1,\\
\label{eq:diff}
\hspace*{-0.8cm}H=1/2:&&\gamma_Y^2(f,T)\sim1+\frac{1}{4}\left(1+\frac{
\sigma_e^2\tau_0^2}{2D}f^2\right)^{-2},\\
\label{eq:sup}
\hspace*{-0.8cm}H>1/2:&&\gamma_Y^2(f,T)\sim1+\left(1+\frac{\sigma_e^2\tau_
0^2}{D}\frac{f^2}{T^{2H-1}}\right)^{-2}.
\end{eqnarray}
\end{subequations}
Thus, $\gamma_Y$ is completely independent of the static noise for 
subdiffusion, while for normal and superdiffusion correction terms enter. 
In these, the limit of zero frequency leads us back to the values in absence of 
noise, that is $\gamma^2_Y (f=0) = 2$ (see Supplementary Note 2
for more details).
Note that for superdiffusion  the limit of long measurement times also leads 
to the noise-independent value of $\gamma_Y$.

\subsection{Analysis of simulation data}

\begin{figure*}
\includegraphics[width=16cm]{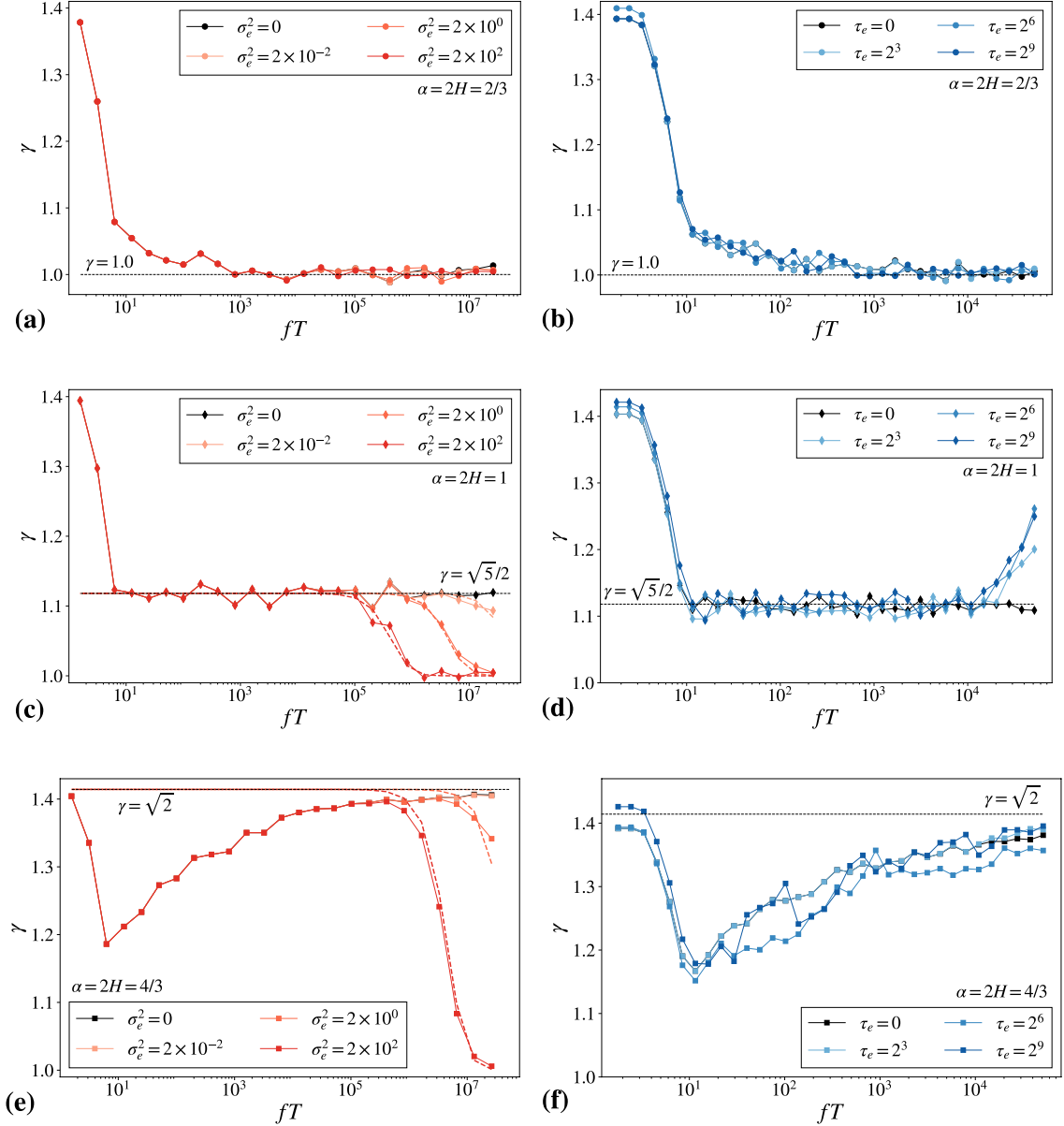}
\caption{{One-dimensional fractional Brownian motion (FBM) trajectories.
We show the coefficient of variation from Monte Carlo simulations of
one-dimensional FBM trajectories in the presence of localisation noise.}
We set the generalised diffusion coefficient to $D=1$ and the Hurst exponent
to (a)--(b) $H=1/3$ (subdiffusion), (c)--(d) $H=1/2$(normal diffusion) and 
(e)--(f) $H=2/3$ (superdiffusion).  Panels (a), (c) and
(e): $n=10^4$ realisations consisting of $N=2^{23}$ discrete time steps with
$\Delta t=1$ from the joint process defined in \eqref{Y_t}, for static error
only. The dashed lines represent the expected high-frequency asymptotic trend
reported in Eqs.~\eqref{eq:sub} to \eqref{eq:sup} for the different values of
$\sigma_e$ and $\tau_0=1$. Panels (b), (d) and (f): $n=10^4$ realisations consisting
of $N=2^{14}$ final steps with $\Delta t=1$ obtained for pure FBM (black) and
in the presence of dynamic error with $\tau_e$.  The dashed lines represent
the limiting value at high-frequencies for pure FBM.  Note that $\gamma$
is reported as a function of $f T$, thus the limiting values obtained here
for high-frequencies are also valid for the case of fixed $f$ and large $T$.}
\label{img1}
\end{figure*}

\begin{figure*}
\includegraphics[width=16cm]{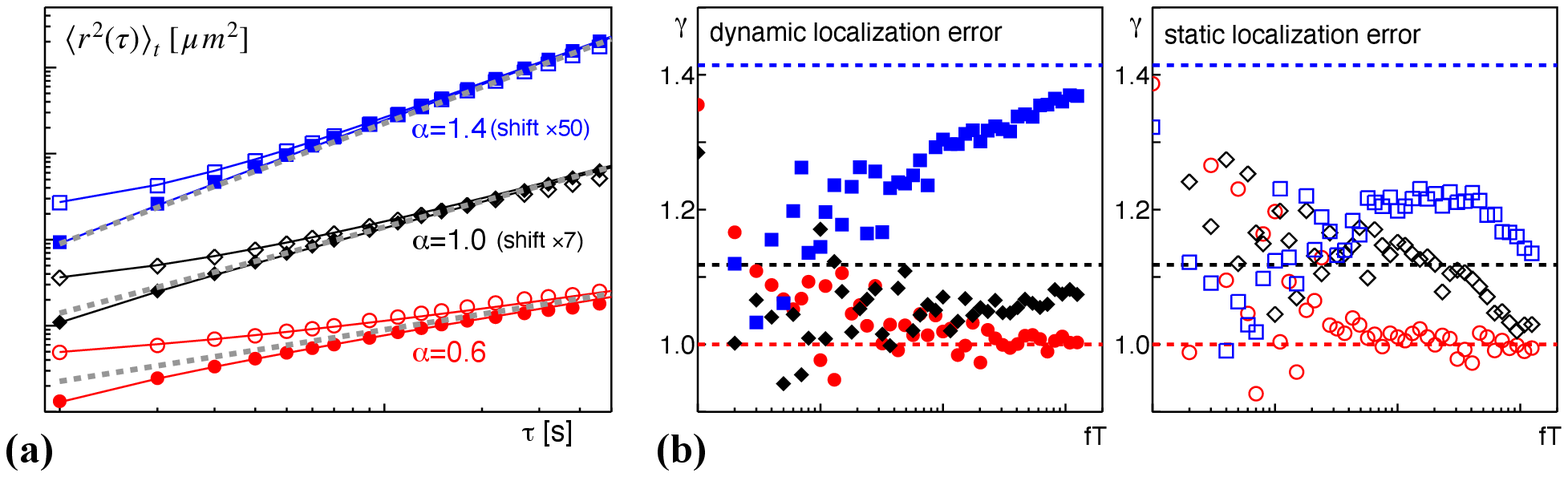}
\caption{{Two-dimensional fractional Brownian motion (FBM) trajectories.
We show the time-averaged mean squared displacements (TA-MSDs) and the
coefficient of variation from Monte Carlo simulations of two-dimensional 
FBM trajectories in the presence of localisation noise.}
(a) Representative time-averaged mean squared displacements
(TA-MSDs) of individual two-dimensional FBM trajectories created with
$H=0.3$ (red circles), $H=0.5$ (black diamonds), and $H=0.7$ (blue
squares) follow the anticipated power-law scaling (grey dashed lines)
for sufficiently long lag times $\tau$. Ensemble averages of TA-MSDs
are superimposed as coloured full lines. For short lag times,
significant deviations due to a dominant dynamic localisation error
($n_p=900$, filled symbols) or a dominant static localisation error
($n_p=50$, open symbols) are visible. These may significantly perturb
the extraction of the TA-MSD's scaling exponent, $\alpha=2H$. (b) The
coefficient of variation $\gamma$, for the case of a dominant dynamic
localisation error (filled symbols) converges towards the predicted
values $\gamma=1,\sqrt{5}/2$ and $\sqrt{2}$ (highlighted by the
coloured dashed lines). While the subdiffusive case (red) converges
rapidly, normal diffusion and superdiffusion (black and blue,
respectively) may need longer trajectories to eventually reach the
predicted value. In contrast, for a dominant static localisation error
(open symbols) only the subdiffusive case (red) is in agreement with the
predicted value $\gamma=1$ whereas normal diffusion and superdiffusion
(black and blue) are very sensitive to this perturbation.}
\label{img2}
\end{figure*}

We start our discussion with results from analytical predictions and simulations
({see "Analytical predictions" and "Simulations" subsections in 
Methods}). Results from 1D simulations are shown in Fig.~\ref{img1}. 
The main goal of this analysis is to elucidate the separate contributions of the two 
localisation errors, i.e. static and dynamic, in the study of the coefficient 
of variation $\gamma$. Results from simulations in 2D are reported in 
Fig.~\ref{img2}. The latter are obtained following a procedure that imitates 
a real experiment ({see "Simulations" subsection in
Methods} for more details) and thus provide more realistic results to be 
compared with the ones showed below from experiments.

In Fig.~\ref{img1}, panels (a)-(b) we immediately observe that the limiting
value of $\gamma$ in the case of subdiffusive FBM ($H=1/3$) is not affected
by any of the two measurement noises. In the presence of static error only
this result was proved also analytically in Eq.~\eqref{eq:sub}.

For the normal diffusive case ($H=1/2$), in Fig.~\ref{img1}, 
panels (c)-(d), the situation differs distinctly from the subdiffusive case. 
Namely, the limiting value of $\gamma$ is affected by both types of 
measurement noise, resulting in two opposite effects. On the one hand, 
the static localisation error leads to a drop in the value of $\gamma$ 
(Fig.~\ref{img1}, panel (c)). In particular, from the analytical 
result Eq.~\eqref{eq:diff} one can see that when the ratio $\sigma^2_e/D$ 
grows, $\gamma$ decreases, approaching the new limiting value 
$\gamma=1$ at very high frequencies. On the other hand, from our 
numerical study in Fig.~\ref{img1}, panel (d) one can see 
that the dynamic localisation error causes an increase in the limiting 
value of $\gamma$. The greater $\tau_e$ the smaller the value of $fT$ at
which we observe a deviation from the pure value $\gamma=\sqrt{5}/2$.

For superdiffusive FBM, the results are reported in Fig.~\ref{img1}, 
panels (e)-(f), showing that the limiting value of $\gamma$ is mostly
affected by the presence of static localisation errors. The latter, similarly
to the diffusive case, leads to a drop in the limiting value of $\gamma$. In
particular, if we look at the explicit analytic expression reported in
Eq.~\eqref{eq:sup} we see that upon increasing the ratio $\sigma^2_e/D$
the coefficient of variation drops, until it reaches the new limiting value
$\gamma=1$, as for BM ($\alpha=1$). Moreover, in this case the trajectory
length $T$ also turns out to be a key parameter in determining the final
trend of $\gamma$: for very long trajectories the presence of static
measurement noise becomes less and less influential. For the dynamic
error instead we see from Fig.~\ref{img1}, panel (f) that, regardless 
of the value of $\tau_e$, the trend of $\gamma$ does not present any 
relevant change.

In general, we observe that localisation errors affect the trend of 
$\gamma$ for normal diffusion and superdiffusive FBM only. For both cases
($\alpha=1$ and $\alpha>1$), when the variance $\sigma_e$ of the static
error increases, the drop of $\gamma$ to the ultimate value 1 occurs at 
progressively lower values of $fT$. Instead, when changing the key 
parameter of the dynamic error ($\tau_e$), the trend of $\gamma$ 
is affected only in the case of normal diffusion, showing an increase in 
its limiting value.

The analysis of numerically obtained FBM data in 2D with different localisation
errors is reported in Fig.~\ref{img2}. As can be seen, these results agree with
the discussion for the case of the 1D simulations. On the one hand, in the case
of subdiffusive FBM the coefficient of variation is insensitive to any of the
localisation errors, always converging to the prediction $\gamma=1$. Therefore,
this measure is a very robust mean to explore whether tracking data indeed
show a subdiffusive behaviour. On the other hand, diffusive and superdiffusive
FBM data do not appear to be very sensitive to the presence of dynamic noise,
but their results change dramatically when static noise is present, causing
a clear drop in the value of $\gamma$ at high frequencies.

\subsection{Analysis of experimental data}

\begin{figure}
\includegraphics[width=8cm]{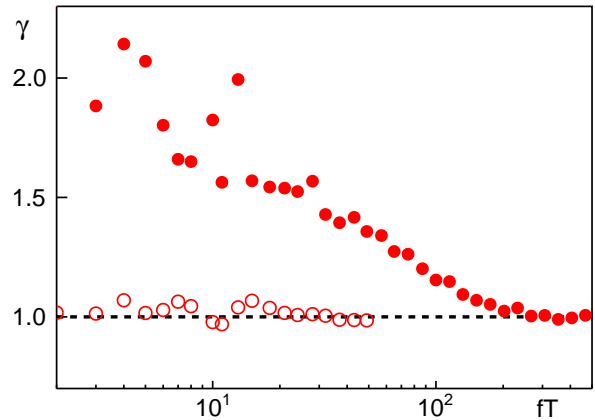}
\caption{{Coefficient of variation obtained from experimental subdiffusive 
trajectories with a marked anti-persistence.} Results from two 
datasets are shown: experimental tracking of p-granules in {\em C. elegans} 
embryos (open red circles) and telomeres in the nucleus of mammalian 
culture cells (filled red circles). They both comply with the expectation 
for subdiffusive data that $\gamma$ converges to unity (indicated by the 
black dashed line). In both cases, the MSDs contained a non-negligible 
static localisation error.}
\label{img3}
\end{figure}

Analytical and numerical results are nicely confirmed by the analysis of
experimental data. Fig.~\ref{img3} shows results from subdiffusive tracking
data on (i) telomeres in the nucleus of mammalian culture cells \cite{SW2017},
and (ii) p-granules in the cytoplasm of {\em C. elegans embryos} \cite{BW2021},
all distinctly displaying a convergence to $\gamma=1$, fully consistent with
our predictions.

\begin{figure}
\includegraphics[width=8cm]{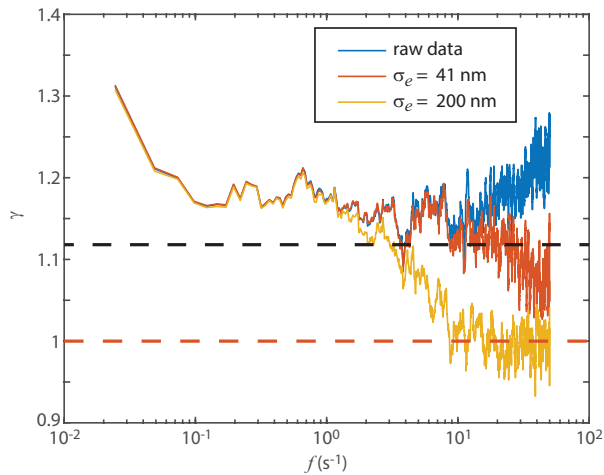}
\caption{{Coefficient of variation obtained from diffusive data of beads
in water.} The base-line is marked in blue. Static localisation errors 
are artificially added to the data to obtain the other two curves. The red 
and black dashed lines indicate $\gamma=1$ and $\sqrt{5}/2$, respectively, 
which are the values to which $\gamma$ converges in the case of 
subdiffusion and normal diffusion, when no error is present.}
\label{img4}
\end{figure}

Experimental data obtained by tracking the motion of beads in water, reported
in Fig.~\ref{img4}, show some nice results for the normal-diffusion regime.
For the pure experimental data (base-line) we can see an increase in the value of
$\gamma$ at high frequencies. We showed that this deviation from the expected
value is due to the effect of the dynamic error. Moreover, if in a post-analysis
we artificially increase the static localisation error, we observe that the
value of $\gamma$ starts decreasing upon increase of $\sigma_e$, approaching
the new limiting value $\gamma=1$ (red dashed line in Fig.~\ref{img4}). These
observations are fully in line with our predictions.

\begin{figure*}
\includegraphics[width=16cm]{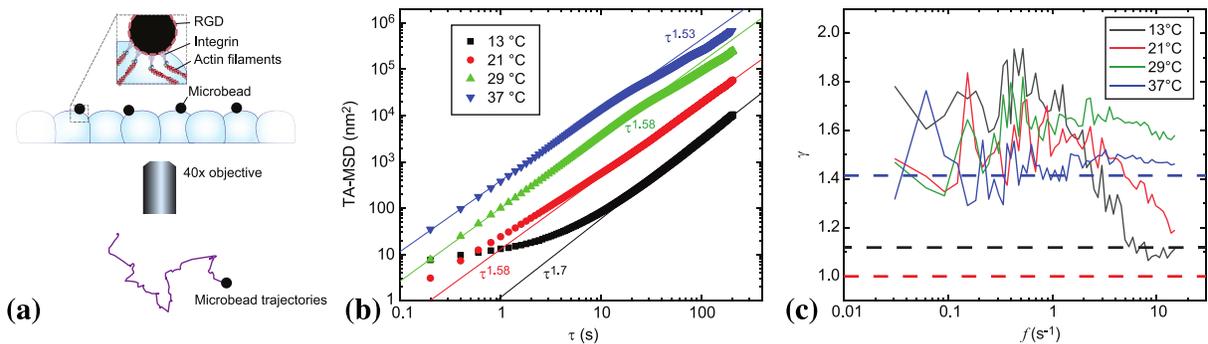}
\caption{{Cytoskeleton fluctuations of living cells exhibit superdiffusive
behaviour.} (a) Sketch of the experimental system for microbeads
attached to the surface of epithelial cells (see the "Experiments"
subsection in the Methods for more details) together with a 
representative experimental trajectory. (b) MSD at varying temperatures. 
The straight lines show that the dynamics are superdiffusive with 
$2H$ between $1.5$ and $1.7$, depending on temperature. 
(c) Coefficient of variation of the power spectrum of the motion of 
the surface-bound microbeads. The dashed lines indicate
$\gamma=1$, $\sqrt{5}/2$, and $\sqrt{2}$, the predicted limiting values of
$\gamma$ for the different diffusion regimes.}
\label{img5}
\end{figure*}

We now move to the analysis of superdiffusive experimental data. 
Cytoskeleton fluctuations were measured by attaching RGD-coated 
beads to the surface of human alveolar epithelial cells (see sketch in 
Fig.~\ref{img5}, panel (a)). The beads were connected to the 
actin cytoskeleton and rearranged by internal molecular motors. 
Given that molecular motor activity depends on temperature, the data 
show a marked superdiffusive behaviour at high temperatures 
($\geq$ 29$^{\circ}$C, green triangles and blue circles in Fig.~\ref{img5},
panel (b)). Interestingly, at low temperatures ($\leq$ 21 $^{\circ}$C) 
the data show that cytoskeleton fluctuations transition from a subdiffusive 
behaviour at short timescales to a superdiffusive behaviour at longer 
timescales (black squares in Fig.~\ref{img5}, panel (b)). This crossover 
can be explained by the Arrhenius dependence of the activity of molecular 
motors on temperature \cite{preFR}. Using the coefficient of variation $\gamma$ 
to assess the diffusion regime, we observe that cytoskeletal fluctuations 
obtained at high temperatures (29 and 37 $^{\circ}$ C) display a robust 
superdiffusive behaviour at all different timescales. The subdiffusive
to superdiffusive crossover at 13 and 21 $^{\circ}$ C is well captured
as $\gamma$ approaches $1$ for frequencies $>20$ rad/s (Fig.~\ref{img5}, 
panel (c)). At longer timescales the value of $\gamma$ indicates 
superdiffusive behaviour, showing that at low temperatures, molecular 
motors are active at longer timescales. From the analysis of this dataset 
we can draw two conclusions. First, we see that there is no statistically 
relevant effect due to localisation errors. Second, we observe that the 
value of $\gamma$ is very sensitive to crossovers between different 
diffusive regimes, confirming that this quantity is useful for the statistical 
analysis of experimental data.

\begin{figure*}
\includegraphics[width=10cm]{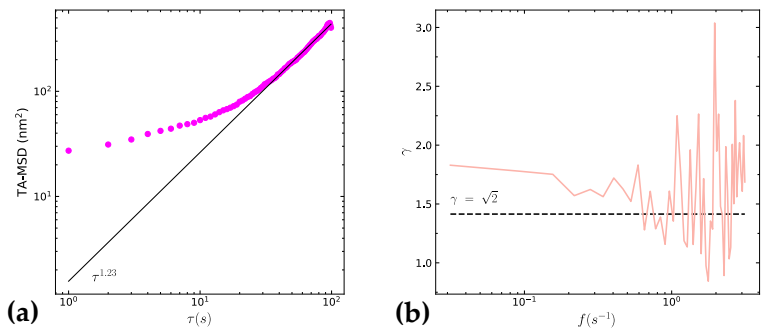}
\caption{{Mean squared displacement (MSD) and coefficient of variation for 
superdiffusive nanoparticles in hMSCs.} (a) MSD showing a superdiffusive 
regime with large deviations due to the presence of the static localisation error. 
(b) The coefficient of variation confirms the superidffusive trend. }
\label{img6}
\end{figure*}

Finally, we consider the diffusion of nanoparticles in the cytoplasm of human
Mesenchymal stem cells. In Fig.~\ref{img6}, panel (a) we report the MSD
showing a mildly superdiffusive ($\alpha=1.23$) regime, in agreement 
with the fact that the nanoparticles are embedded into the active intracellular
environment. However, the MSD trend is strongly affected by the presence 
of the static localisation error, making the fit less reliable. By calculating the
coefficient of variation, reported in Fig.~\ref{img6}, panel (b), we clearly confirm 
the superdiffusive trend, as $\gamma$ converges nicely to the analytical 
value for superdiffusion regime, that is $\sqrt{2}$, without displaying large 
deviations due to the static error.

\section{Conclusions}

Departures from the Brownian behaviour of diffusive processes are
observed in a wide variety of systems of practical interest across many
disciplines, and these phenomena call for explanations and understanding
of the underlying physical processes and microscopic mechanisms. Without
such a knowledge one cannot fully comprehend and reach a full picture
of the phenomena. Concurrently, a conclusion that the dynamics is indeed
anomalous relies on proper data treatment, the size of statistical samples,
blurring measurement errors and errors incurred by the fitting procedures,
or due to transients obscuring the true scaling exponents. It is therefore
indispensable to have at hand robust methods allowing one to reach justified,
sound conclusions on the dynamics.

In this work we test the coefficient of variation $\gamma$ of the
single-trajectory power spectral density and show that it represents a 
valuable method towards a robust criterion for anomalous diffusion.
We combine analytical, numerical and experimental 
studies of diffusive dynamics in very diverse systems, to demonstrate how 
the values of $\gamma$ for trajectories belonging to FBM are affected by 
the presence of localisation errors. Within such a combined effort, numerical 
simulations which imitate real experiments performed here serve us to 
elucidate the relative contributions of static and dynamic measurement 
noise in a controllable way, and therefore to prove our theoretical results 
and concepts. A further comparison with experimental data obtained for 
rather diverse systems permits us to verify the predicted trends and thus 
obtain a fully comprehensive picture.

Apart from the case $f\to 0$ in which the coefficient of variation $\gamma(f,T)$ 
converges to the universal value $\sqrt{2}$, we found that our results vary 
depending on the specific diffusive regime and on the kind of localisation error.
In particular, the coefficient of variation represents a very robust way to define 
whether tracking data show a subdiffusive behaviour. This can be of decisive 
help in data analysis, in particular, when the deviation of the anomalous exponent
from unity is small and thus the fitting of the MSD can produce misleading
results. Conversely, diffusive and superdiffusive data are more sensitive
to measurement noise. In these regimes the value of $\gamma$ displays a
clear drop at high frequencies in the presence of a static localisation error,
which is corroborated by both numerical and experimental data,
while the effect of a dynamic error appears to be relevant for the normal-diffusive
regime only, causing an increase in the limiting value of $\gamma$ at high
frequencies. As well, this theoretical prediction is confirmed by our numerical 
and experimental results. Thus, in the case of superdiffusion the criterion 
remains very robust in the presence of dynamic error while in general it is not 
when static localisation error is present. Nevertheless, the analytical expression 
obtained for the correction term still allows us to control the effect of the 
measurement error in this case. Finally, the criterion appears to be the least 
robust for normal diffusion, which is affected by both static and dynamic 
localisation errors.
Arguably, the observed trends in $\gamma$ can be used also to post-process 
the data by adding static and dynamic error to explore the limiting values 
of $\gamma$ as function of noise intensity. 

The application of our method to diverse experimental systems presented in our
paper evidences that it provides a very robust tool for the analysis of the
anomalous character of random motion. In conjunction with the MSD analysis,
it will permit to make conclusive statements on the actual departures from
standard diffusive motion, even in the presence of unavoidable errors in
experimental measurements, and thus to point towards the necessity 
to understand the actual underlying physical mechanisms. Apart from a 
stand-alone criterion our power spectral method will also represent an 
important ingredient in decision trees and feature-based neural 
network approaches.

\section{Methods}

\subsection{Simulation methods}

We perform simulations of FBM trajectories in 1D and 2D in order to complement
and support our analytical predictions as well as the analysis of experimental
data. The simulations are performed in Python for the 1D case and in Matlab
for the 2D case.

\subsubsection{1D case}

FBM trajectories of length $T=N\times\Delta t$, where $N$ corresponds to the
total number of time steps and $\Delta t$ to the discretisation time step,
are generated for different values of the Hurst index $H$ to explore all 3
regimes via the {\em fbm} package in Python. In order to include the dynamic
error, given a fixed exposure time $\tau_e$ corresponding to $n_e=\tau_e/
\Delta t$ time steps, we first simulate longer FBM trajectories with a total
number $N_e=N\times n_e$ of time steps. Then, from each of these trajectories
a new trajectory is obtained, whose points are given by the average over $n_e$
subpoints of the original trajectory, i.e., $\overline{X}(t_i)=\frac{1}{n_e}
\sum_{j=0}^{n_e-1} X(t_i-j\Delta t)$. As far as the static error is concerned
we use its definition (\ref{stat_err}) and generate stationary OU trajectories
with relaxation time $\tau_0=\Delta t$ and varying noise amplitude $\sigma_e$.

\subsubsection{2D case}

An ensemble of $M=100$ two-dimensional FBM trajectories, each with 
$N=2500$ positions covering a total time $T=N \times\Delta t$ with 
$\Delta t=0.1$~s was created with the Matlab routine {\em wfbm} for the Hurst 
coefficients $H=0.3$, $0.5$, and $0.7$, hence exploring FBM from the 
subdiffusive to the superdiffusive regimes. Mean step sizes within $\Delta t$ 
along each coordinate were chosen to be $0.01~\mu$m.  To account for the 
dynamic and static localisation errors, each time step was subdivided into 
$n_e=10$ substeps during each of which $n_p$ positions according to a
diffraction-limited Gaussian PSF, centred at the particle position at that
time, were calculated. To this end, initial FBM tracks had a length $i=1,
\ldots,Nn_e$ and at each substep location $x_i$, a total of $j=1,\ldots,n_p$
positions $x_{i,j}=x_i+\xi_j$ were produced, where $\xi_j$ are Gaussian
random numbers with zero mean and standard deviation $\sigma=0.220~\mu$m.
From this, track positions were determined via
\begin{equation}
X_k=X(k\Delta t)=\frac{1}{n_e n_p}\sum_{i=(k-1)n_e+1}^{kn_e}
\sum_{j=1}^{n_p} x_{i,j}.
\end{equation}
For $n_p<100$, this procedure creates a dominant static localisation error
whereas for $n_p\gg100$ only a dynamic localisation error is seen.

\subsection{Experimental systems}

To assess the applicability to experimental data of the coefficient of
variation analysis in the presence of localisation errors, we analysed
experimental data displaying different diffusive regimes. In what follows
we report a short description of the experimental systems.

\subsubsection{Subdiffusive data}

{\em Experimental tracking data for p-granules in C. elegans embryos.}
The data were obtained and analysed as described in \cite{BW2021}. 
As shown before, p-granule trajectories have a noticeable static localisation 
error that perturbs, e.g., the velocity autocorrelation function. Here, only
trajectories with $N=100$ positions at a time increment $\Delta t=210$~ms
with a subdiffusive TA-MSD scaling were considered. Trajectories with
scaling exponents $\alpha\in[0.7,0.9]$, $\alpha\in[0.5,0.7]$, and
$\alpha\in[0.3,0.5]$ were grouped into three distinct sets. For each,
$\gamma$ was calculated as a function of $fT$, and the average of these
was used for Fig.~\ref{img3} to soften fluctuations induced by the fairly
short trajectories.

{\em Trajectories for telomeres in the nucleus of mammalian culture cells.} 
The data were obtained similar to our previous work \cite{SW2017}: U2OS 
cells (DSMZ Cat\# ACC-785, RRID:CVCL\_0042) were cultured as described 
and telomeres were highlighted by transient transfection with a plasmid for 
GFP-tagged TRF-2 (24~h prior to microscopy, using Lipofectamine3000 
according to the manufacturer's protocol). For live-cell microscopy, cells 
were plated 24~h prior to transfection in 4-well $\mu$-slide microscopy chambers; 
15~min prior to imaging, the medium was changed to MEM without 
phenol red supplemented with 5\% FCS and 5\% HEPES. 
Imaging was performed with a customised spinning-disk confocal microscope 
based on a DMI~4000 stand (Leica Microsystems, Germany) with a custom-made 
incubation chamber, a CSU-X1 spinning-disk unit (Yokogawa Microsystems, Japan), 
an HC PL APO 100x/1.40NA oil immersion objective, and an Evolve 512 EMCCD
camera (Photometrics, USA). Samples were illuminated at 488~nm
and fluorescence was detected in the range 500-550~nm. Trajectories were
recorded at a time increment $\Delta t=110$~ms and only trajectories with
at least $N=1000$ positions were retained (and trimmed to the same
length if trajectories were longer). In line  with our previous finding,
trajectories featured an anti-persistent subdiffusion with an average
scaling exponent of $\alpha\approx 0.55$.

In both cases, the generalised diffusion coefficients of individual
trajectories varied log-normally. To soften the influence of this locus-
and particle-dependent prefactor in the PSD analysis, each trajectory was
normalised in each coordinate by its respective root-mean-squared step
taken within $\Delta t$ before calculating $\gamma$.

\subsubsection{Diffusive data}

We analysed the motion of 1.2 $\mu\rm{m}$-sized polystyrene beads in aqueous
solution as representative experimental data for Brownian motion. Namely, we
suspended the beads in phosphate-buffered saline with 1\% bovine serum
albumin and 0.05\% Tween 20 to avoid aggregation, and introduced the
suspension into a flow cell chamber. Subsequently, the flow cell chamber was
sealed for imaging. The beads were imaged at 100 frames per second in an
inverted microscope with a 40x objective (Olympus PlanApo, N.A. 0.95) and
a sCMOS camera (Andor Zyla 4.2). Bead tracking in the plane was performed
in LabView using a cross-correlation based tracking algorithm. A set of 150
trajectories, each consisting of 4096 frames, was used in the analysis.

\subsubsection{Superdiffusive data}

{\em Cytoskeleton fluctuations of living cells at different temperatures \cite{preFR}.} 
Cytoskeleton fluctuations were measured by tracking the trajectory of 
refractive microbeads attached to the surface of human alveolar epithelial cells 
(A549). The microbeads were previously coated with Arginine-Glycine-Aspartic 
acid (RGD) containing peptide (Peptide 2000, Integra life Sciences, San Diego, 
CA) to link the probe to the actin cytoskeleton through integrin membrane 
receptors. The movement of these beads was sensitive to a wide range of 
cytoskeleton manipulations including actin polymerisation/depolymerisation 
drugs, actomyosin relaxation, cell stretching, temperature changes, and 
ATP depletion \cite{NM_Bursac} indicating that spatial fluctuations of 
microbeads linked to integrin membrane receptors reflect intrinsic 
cytoskeleton dynamics.

Microbead positions were tracked at 40x magnification using an inverted
microscope (TE-2000E, Nikon, Japan) equipped with a charge coupled device
(CCD) camera (Orca, Hamamatsu, Japan). The spontaneous microbead movement was
tracked for 200-400s at a sampling rate of 5 Hz. The position of the microbead
was determined with subpixel resolution by computing the microbead centroid
with an intensity-weighted average algorithm implemented with a custom-made
software (LABVIEW, National Instruments, U.S.). Data were corrected for
the drift of the stage of the microscope, which was computed as the average
change in the position of all microbeads within the field of view.

The temperature dependence of cytoskeleton fluctuations was measured by
heating or cooling the microscope stage with a microincubator system
(HCMIS MicroIncubator System, ALA Science, Westbury, NY) and closed-loop
control. The sample temperature was measured with a negative temperature
coefficient thermistor (332 Temperature Controller, Lakeshore, Westerville,
OH). Measurements were taken in $n=6$ wells ($\sim$ 20 microbead/well)
per temperature.

Human alveolar epithelial cells (A549) (cell line CCL-185 ATCC, Manassas,
VA) were cultured in RPMI 1640 medium supplemented with 1 mM L-glutamine,
100 U/ml penicillin, 100 mg/ml streptomycin and 2 $\mu$g/ml amphotericin
B (all from GIBCO, Gaithersburg, MD), 10$\%$ inactivated fetal calf serum
(Biological Industries, Kibbutz Beit Haemek, Israel), and buffered with HEPES
(Sigma, St. Louis, MO). One day before experiments cells were harvested
with a brief exposure to trypsin-EDTA (Sigma) and plated (900 cells/mm$^2$)
on collagen-coated wells.

{\em Diffusion of nanoparticles in the cytoplasm of human
Mesenchymal stem cells.} Here we tracked yellow-green fluorospheres 
of size 100~nm injected carefully in the cytoplasm of human mesenchymal 
stem cells (hMSCs). The fluorospheres (FluoSpheres\textsuperscript{TM} 
ThermoFisher, Cat.~No: F8803) were negatively charged (carboxylated-modified) 
polystyrene beads that are suitable for intracellular tracking. Samples 
were prepared by diluting the suspension to 2 $\text{mg}/\text{ml}$ 
concentration after 20 minutes sonication of the stock solution to ensure 
even dispersion of the particles in the solution. 
The diluted solution subsequently was vortexed
for two minutes for optimised mixing and then loaded via a microloader
pipette (Eppendorf) in manufactured glass capillaries appropriate for
microinjection (Femtotips\textsuperscript{\textregistered} Eppendorf Cat.~Nr.:
5242952008). Their microinjection was executed at room temperature using
a micromanipulation system (Eppendorf) at controlled pressure. Immediately
after the injection, imaging was performed 100 s at room temperature with an
Olympus IX81 inverted microscope using a Olympus UPLSAPO40X/0.95 Objective
and a Hamamatsu Orca-2 camera. At each frame of the resulting video files
the particles were identified and their movement were tracked using a publicly
available Python package which also provides tools to spot the candidate
features based on high intensity matches, filtering and different type of
corrections such as drift correction \cite{trackpy}.

\begin{acknowledgments}
RM acknowledges the German Science Foundation (DFG, grant No. ME 1535/12-2)
for support. MW and RB acknowledge financial support by VolkswagenStiftung
(Az 92738) and by the Elite Network of Bavaria (Study Program Biological
Physics). CS acknowledges the VolkswagenStiftung (Az 96733).
\end{acknowledgments}

\end{document}